**Applications of Post-quantum Cryptography**

Emils Bagirovs, Grigory Provodin, Tuomo Sipola, Jari Hautamäki
Institute of Information Technology, Jamk University of Applied Sciences, Jyväskylä, Finland
emils.bagirovs@student.jamk.fi
grigory.provodin@student.jamk.fi
tuomo.sipola@jamk.fi
jari.hautamaki@jamk.fi

**Abstract:** With the constantly advancing capabilities of quantum computers, conventional cryptographic systems relying on complex math problems may encounter unforeseen vulnerabilities. Unlike regular computers, which are often deemed cost-ineffective in cryptographic attacks, quantum computers have a significant advantage in calculation speed. This distinction potentially makes currently used algorithms less secure or even completely vulnerable, compelling the exploration of post-quantum cryptography (PQC) as the most reasonable solution to quantum threats. This review aims to provide current information on applications, benefits, and challenges associated with the PQC. The review employs a systematic scoping review with the scope restricted to the years 2022 and 2023; only articles that were published in scientific journals were used in this paper. The review examined the articles on the applications of quantum computing in various spheres. However, the scope of this paper was restricted to the domain of the PQC because most of the analyzed articles featured this field. Subsequently, the paper is analyzing various PQC algorithms, including lattice-based, hash-based, code-based, multivariate polynomial, and isogeny-based cryptography. Each algorithm is being judged based on its potential applications, robustness, and challenges. All the analyzed algorithms are promising for the post-quantum era in such applications as digital signatures, communication channels, and IoT. Moreover, some of the algorithms are already implemented in the spheres of banking transactions, communication, and intellectual property. Meanwhile, despite their potential, these algorithms face serious challenges since they lack standardization, require vast amounts of storage and computation power, and might have unknown vulnerabilities that can be discovered only with years of cryptanalysis. This overview aims to give a basic understanding of the current state of post-quantum cryptography with its applications and challenges. As the world enters the quantum era, this review not only shows the need for strong security methods that can resist quantum attacks but also presents an optimistic outlook on the future of secure communications, guided by advancements in quantum technology. By bridging the gap between theoretical research and practical implementation, this paper aims to inspire further innovation and collaboration in the field.

## 1. Introduction

As quantum computers become more powerful, they bring new risks to the current security systems. Traditional cryptographic methods, like those used in online banking and email encryption, rely on complex math problems that classical computers find hard to solve. However, quantum computers, with their advanced capabilities, can solve these problems much faster, making many of the current security methods weak against quantum attacks (Shaller, Zamir and Nojoumian, 2023; Lei et al., 2023).

Quantum computers employ multiple quantum mechanics phenomena such as superposition, entanglement, non-cloning theorem, and more. Superposition is a cornerstone of the quantum bits (qubits). While classical computers can be only in one state of 0 or 1, quantum computers can be in two states simultaneously. This aspect of superposition enables a dramatic increase in the speed of quantum computers. Entanglement is a unique state where two particles are interconnected, meaning that these particles when observed can provide information about each other, despite the distance between them. In quantum computation, this phenomenon is mostly used in optimization problems, secure communications, and quantum internet. No-cloning theorem states that the perfect cloning of a quantum state is impossible. The means of information as we know it today will be very

Funded by the European Union.

Funded by the European Union. Views and opinions expressed are however those of the author(s) only and do not necessarily reflect those of the European Union or the European Commission. Neither the European Union nor the granting authority can be held responsible for them.

This is the author accepted manuscript version. The original article appeared as: Bagirovs, E., Provodin, G., Sipola, T., & Hautamäki, J. (2024). Applications of Post-quantum Cryptography. In *Proceedings of the 23rd European Conference on Cyber Warfare and Security (ECCWS)*. Academic Conferences International. https://doi.org/10.34190/eccws.23.1.2247

different with the employment of quantum computers (Belkhir, Benkaouha and Benkhelifa, 2022; Sridhar, Ashwini and Tabassum, 2023; Zhao and Zheng, 2001).

Post-quantum cryptography (PQC) should solve the challenge of the cryptography created by quantum computers. It uses new kinds of math problems that even quantum computers find hard to solve, positioning PQC as a strong candidate for securing our data against future quantum threats (Chawla and Mehra, 2023; Verchyk and Sepúlveda, 2023). However, PQC's practicality is as important as its strength. This means PQC methods should function efficiently on current computers and networks, requiring minimal memory or processing power, crucial for devices like smartphones (Kumar, 2022).

Despite its potential, PQC is still nascent and faces its challenges. These new cryptographic methods haven't undergone as extensive testing as current methods, which have been trusted for many years. PQC, being newer, necessitates more time and testing to gain comparable trust levels. Moreover, transitioning from current methods to PQC is a significant undertaking, involving the complex and costly updating of numerous systems (Yalamuri, Honnavalli and Eswaran, 2022).

This article explores how PQC works, its benefits, and the challenges in applying it to real-world situations. We will also delve into some of the leading PQC methods being considered today, aiming to provide a clear understanding of how post-quantum cryptography can safeguard our data in the quantum computing era.

After an analysis of the quantum computing advancements two research questions were defined. Later, after the evaluation of the obtained sources third research question was defined.

**RQ1:** *What applications are there for quantum computing? And how can they be categorized?*
This question aims to identify various practical applications of quantum computing and to structure the obtained applications.

**RQ2**: *What is the current state of post-quantum cryptography: progress, use cases, and challenges?*
This question aims to identify the latest progress, use-cases, and challenges of the current post-quantum cryptography.

The article is outlined as follows: Section 1 introduces the topic and sets the stage for the discussion. Section 2 outlines the methodology, detailing the criteria for quantum computer application data collection. Section 3 is dedicated to exploring post-quantum cryptography, its types, and applications. Section 4 presents various use cases, while Section 5 discusses the implications and challenges faced in the field. The paper concludes with Section 6, looking towards the future of cryptography in the quantum computing era.

## 2. Methodology

We conducted a scoping literature review that explores quantum computing advancements (Peters et al., 2020). This chapter outlines databases that were used for data collection, inclusion and exclusion criteria that were applied during the review and an overview of the domains that were identified during the review. The paper follows the PRISMA checklist, and its methodology is supported by a PRISMA flow diagram presented in *Figure 1* that shows an overview of the scoping process (Page et al., 2021).

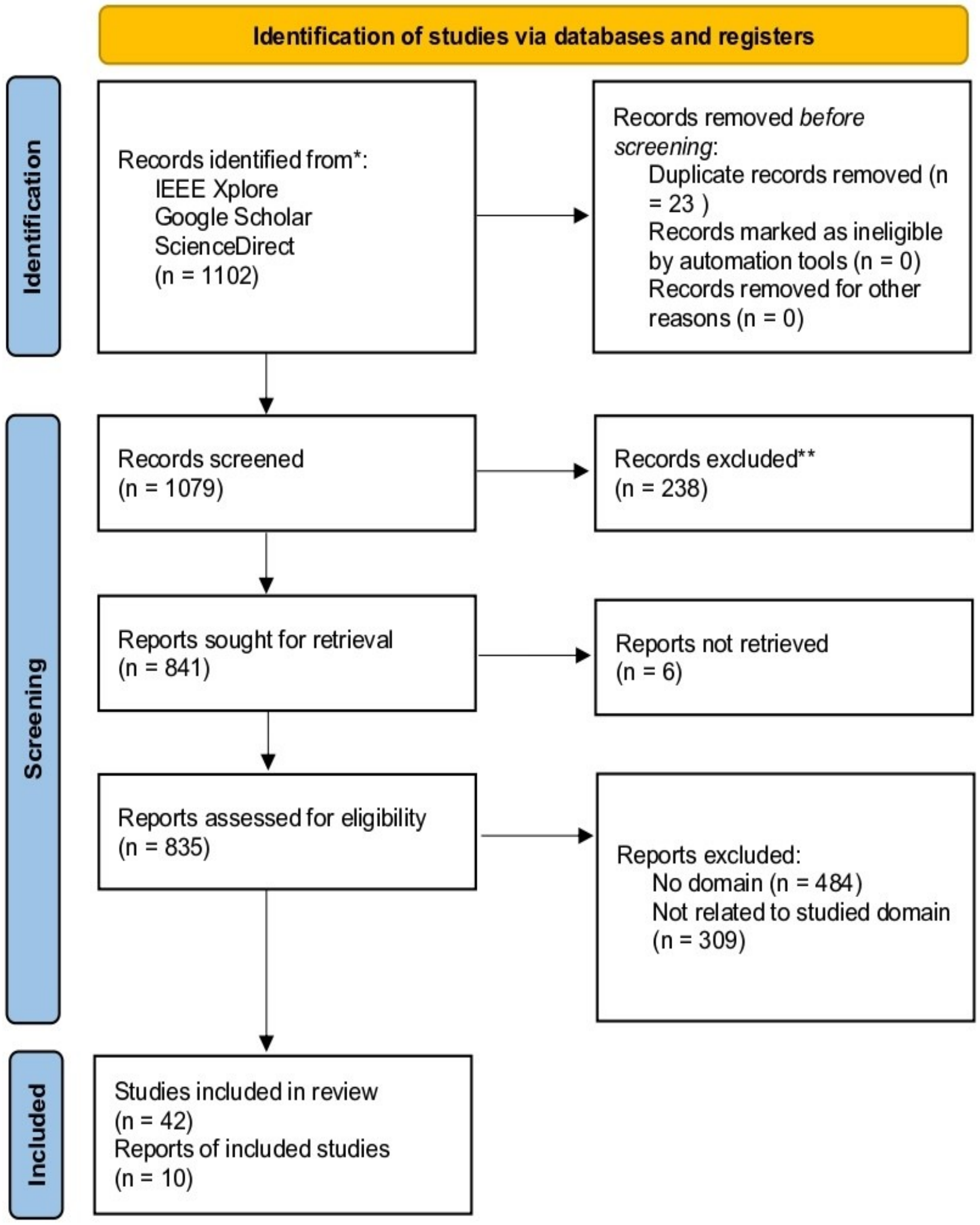

**Figure 1:** PRISMA Flow Diagram

### 2.1 Data collection

This subsection covers means that were applied during data collection: search phrases, databases, inclusion and exclusion criteria, and screening criteria.

#### Search phrases

Based on the research question RQ1 two search phrases were compiled and used across all databases. The two search strings are as follows:

- "Quantum computing applications"
- "Quantum computing use cases"

#### Used databases

For this research three reputable and popular databases were used: IEEE Xplore, which offered a variety of engineering articles; Google Scholar, which provided a range of multidisciplinary articles from various sources; and ScienceDirect, which provided a collection of scientific articles. Our search was limited to the relevant databases that we were able to access through our institution.

**Inclusion and exclusion criteria**

Inclusion and exclusion of the articles were conducted in three stages, where during the first stage only title, abstract, and results sections were considered. As for the second stage, the whole content of the papers was analyzed and evaluated. During the third stage, 42 articles that were included in the review were analyzed based on the scope of the paper. This study focuses on the latest articles, as the progress of the field has accelerated during the last years. Therefore, the selection of the searched papers was restricted to those published between 2022 and 2023. The following criteria and their justifications are outlined below.

**Inclusion criteria**

- Articles are written in English (Aligns with the language of the research and researchers).
- Articles are written between 2022 and 2023 (Ensures focus on recent developments).
- Articles are published in scientific journals (Ensures reliability of the sources).
- Articles are related to quantum computing or/and search phrases (Ensures the correct scope of the review.)

**Screening stage 1:**

- Is the article relevant based on the title, abstract, and results?

**Screening stage 2:**

- Is there an application, and is it usable in practical life?
- Can you identify the topic of the application?

**Screening stage 3:**

- Does this article cover the PQC?
- Can you identify recent developments, algorithms, use-cases, and problems of the PQC?

## 2.2 Domains of Quantum Computer Applications

Out of the 835 retrieved papers 351 had a domain related to applications of quantum computing. The papers and their respective domains can be seen in **Table 1**. Throughout the research, we identified 31 domains of quantum computing applications based on the keywords, headers used, and topics discussed in the papers. The focus of this paper was limited to the domain of Cryptography and Security, in which 42 articles met the inclusion criteria and passed the first two screenings. However, only 10 articles passed the third screening because most of the articles from the Cryptography and Security domain were too niche focusing on the development, calculations, and testing of the variations of PQC algorithms rather than providing an outlook or evaluation.

**Table 1:** Identified domains

| Domain | Paper count |
|---|---|
| Data Science | 7 |
| AI | 13 |
| Algorithms and Applications | 13 |
| Bioinformatics | 4 |
| Challenges and Future Directions | 14 |
| Cloud and Edge Computing | 7 |
| Complex System Optimization | 10 |
| Cryptography and Security | 42 |
| Education and Learning | 1 |
| Engineering and Automated Process Control | 1 |
| Environmental Science | 2 |
| Ethics and Societal Impact | 1 |
| Finance and Economics | 6 |
| Fundamentals and Reviews | 7 |
| Hardware and Architecture | 16 |
| Healthcare | 23 |
| IoT | 7 |
| Machine Learning | 40 |
| Material Science | 23 |
| Nanotechnology | 10 |
| Networks | 12 |
| Optimization Problems | 11 |
| Power System Applications | 5 |
| Renewable Energy | 4 |
| Robotics | 3 |
| Simulation | 12 |
| Smart Cities and Urban Planning | 5 |
| Software and Programming | 13 |
| Space Technology | 4 |
| Specific Industries | 16 |
| Theoretical Physics | 19 |
| **Total** | 351 |

## 3. Post-Quantum Cryptography

Post-quantum cryptography represents a forward-looking approach to cybersecurity, designed to withstand the computational prowess of quantum computers. This segment focuses on the fundamentals of this emerging field, examining specific algorithms and their respective merits and challenges. In addition, **Table 2** is used to represent the ten articles that were included in the final review and shaped this paper.

**Table 2:** Papers included in the review

| **Categories** | **Applications** | **References** |
|---|---|---|
| | **Algorithms** | |
| Lattice-Based | Government communications, financial transactions, and personal data exchanges | Shaller, Zamir and Nojoumian (2023); Yalamuri, Honnavalli and Eswaran (2022); Rewal et al. (2023) |
| Code-Based | Protection of intellectual property | Kumar (2022); Chawla and Mehra (2023) |
| Multivariate Polynomial | Public key cryptosystems | Kumar (2022); Yalamuri, Honnavalli and Eswaran (2022) |
| Hash-Based | Digital signatures | Chawla and Mehra (2023); Yalamuri, Honnavalli and Eswaran (2022); Kumar (2022) |
| Isogeny-Based | Small scale devices | Shaller, Zamir and Nojoumian (2023) |
| | **Future directions** | |
| | New applications of PQC | Aithal (2023) |
| | Requirements for integration of PQC in existing digital infrastructure | Gill et al. (2022) |
| | Necessity in PQC education and training | Rietsche et al. (2022) |
| | **Background** | |
| | Threat from Quantum computers | Lei et al. (2023) |
| | PQC | Verchyk and Sepúlveda (2023) |

Lattice-based cryptography is a prominent example of post-quantum cryptography. It relies on the hardness of lattice problems, which, to date, have no efficient solving algorithm on quantum computers. The advantage of this approach lies in its presumed resistance to quantum attacks, offering a secure alternative to traditional systems (Rewal et al., 2023). However, the downside includes larger key sizes and increased computational overhead, which could be a hindrance in environments with limited resources (Shaller, Zamir and Nojoumian, 2023). Another notable method is hash-based cryptography, which utilizes cryptographic hash functions. These systems are highly efficient and offer strong security assurances against quantum attacks. Their simplicity of design allows for easy integration into existing systems. However, they are not without drawbacks. For instance, hash-based signatures are typically larger than traditional signatures, which can increase data transmission requirements and storage needs (Chawla and Mehra, 2023). Code-based cryptography, drawing security from the hardness of decoding a general linear code, is another viable post-quantum method. Its long-standing presence in the cryptographic community has allowed for substantial analysis and testing. However, like lattice-based systems, code-based cryptography suffers from the requirement of larger key sizes. This characteristic can limit its practicality in systems where bandwidth and storage are constrained (Kumar, 2022). Multivariate polynomial cryptography is yet another approach, which bases its security on the difficulty of solving systems of multivariate polynomials. One of its primary advantages is the potential for smaller key sizes compared to other post-quantum methods. Nonetheless, this advantage comes at the cost of complex key generation and signing processes, which can be computationally intensive (Yalamuri, Honnavalli and Eswaran, 2022). Lastly, isogeny-based cryptography, a relatively newer field, focuses on the computational difficulty of finding isogenies between elliptic curves. This method stands out for its small key sizes and potential efficiency. However, being a newer area of research, it lacks the extensive testing and vetting those older methods have undergone, posing a risk of unknown vulnerabilities (Shaller, Zamir and Nojoumian, 2023).

### 3.1 Categories of Post-Quantum Cryptographic Algorithms

In the field of post-quantum cryptography, various categories of cryptographic algorithms have been developed to counteract the potential threats posed by quantum computers. These categories, each with their unique approach and underlying principles, are pivotal in understanding the landscape of quantum-resistant cryptography (Shaller, Zamir and Nojoumian, 2023; Chawla and Mehra, 2023; Kumar, 2022; Yalamuri, Honnavalli and Eswaran, 2022).

- Lattice-Based Cryptography: This category relies on the complexity of lattice problems, which involve multidimensional grids of points. The security of these algorithms is based on the difficulty of finding the shortest path or closest point in a high-dimensional lattice. Notable for their efficiency and the ability to support advanced cryptographic functions like fully homomorphic encryption, lattice-based algorithms are considered promising for post-quantum cryptography (Shaller, Zamir and Nojoumian, 2023). However, their key sizes and ciphertexts can be relatively large, which might be challenging for systems with limited storage or bandwidth.

- Code-Based Cryptography: Originating from error-correcting codes, this category includes algorithms that are secure due to the difficulty of decoding a general linear code. The most famous example is the McEliece cryptosystem, which has been noted for its fast encryption and decryption processes. While offering strong security, the main drawback of code-based cryptography is the large key sizes, which pose a challenge in terms of storage and transmission efficiency (Chawla and Mehra, 2023).

- Multivariate Polynomial Cryptography: These algorithms are based on the hard problem of solving systems of multivariate polynomials over a finite field. They are known for their fast decryption and potential for small key sizes making them suitable for public key cryptosystems. However, creating secure instances of multivariate polynomial problems can be complex, and some earlier schemes have been broken, raising concerns about their security (Kumar, 2022).

- Hash-Based Cryptography: This type involves cryptographic algorithms that use hash functions. These algorithms are relatively simple and are known for their high speed and security, which rely on the well-studied hardness of finding collisions in hash functions. Hash-based signatures, for example, are a practical application, although they typically have larger signatures than traditional algorithms (Yalamuri, Honnavalli and Eswaran, 2022).

- Isogeny-Based Cryptography: A newer area in post-quantum cryptography, it focuses on the hard problem of finding isogenies between elliptic curves. These algorithms are notable for their small key sizes, making them suitable for systems with limited storage. However, they are less mature than other categories and require more research to fully understand their security and practicality (Shaller, Zamir and Nojoumian, 2023).

Each of these categories represents a different approach to securing cryptographic systems against the threat of quantum computing. The choice of a particular category depends on various factors, including the specific application, the required level of security, and available system resources. As research in this field progresses, these algorithms are continuously evaluated and improved to meet the evolving challenges of cybersecurity in the quantum era (Chawla and Mehra, 2023; Kumar, 2022).

**3.2 Applications of Post-Quantum Cryptography**

The first significant application of Post-Quantum Cryptography (PQC) is in securing communication channels. In the digital age, the exchange of information over the internet, including emails, instant messaging, and online transactions, relies heavily on encryption protocols. Current encryption methods like RSA and ECC are effective against classical computing threats but are vulnerable to quantum attacks. PQC introduces algorithms like lattice-based cryptography, resistant to quantum computing capabilities, making it essential for protecting sensitive information in government communications, financial transactions, and personal data exchanges (Shaller, Zamir and Nojoumian, 2023). Another critical application of PQC is in digital signatures, which are vital for verifying the authenticity of documents and software. Traditional digital signature schemes could be easily compromised with quantum computers. PQC offers solutions like hash-based cryptography, secure against quantum attacks, thereby ensuring the integrity and authenticity of digital signatures in legal documents, software updates, and secure communications (Chawla and Mehra, 2023). Furthermore, the Internet of Things (IoT) is an area where PQC can be transformative. IoT devices, often limited in computational resources, are increasingly involved in critical functions such as healthcare monitoring, smart homes, and industrial automation. Implementing PQC in IoT devices ensures their security in a post-quantum world, protecting them from potential quantum-enabled cyber-attacks (Kumar, 2022). Despite these promising applications, implementing PQC presents challenges. One of the main issues is the increased computational and storage requirements compared to classical cryptography, problematic for devices like IoT devices. Additionally, the relative newness of PQC means a lack of long-term studies and real-world testing, raising concerns about potential vulnerabilities. The application of PQC is crucial in securing the digital world against future quantum threats. Its implementation in secure communications, digital signatures, and IoT devices highlights its importance. However, addressing challenges in computational requirements and the need for further research and testing is necessary to fully realize PQC's potential in these applications.

**3.3 Challenges and Considerations**

One major challenge is the increased computational resources required by many PQC algorithms. Compared to traditional cryptographic methods, PQC often demands more processing power and larger memory spaces, especially for key storage and management. This can be particularly problematic for devices with limited resources, such as smartphones and IoT devices. Implementing PQC in these devices requires careful optimization to balance security needs with available resources (Shaller, Zamir and Nojoumian, 2023).

Another consideration is the current lack of standardization in PQC algorithms. The cryptographic community is still in the process of researching and identifying the most secure and efficient post-quantum algorithms. This ongoing research means that the algorithms have not yet undergone the extensive testing and validation that classical cryptographic algorithms have. There is a need for broad agreement on standards to ensure compatibility and security across different platforms and applications (Chawla and Mehra, 2023).

The integration of PQC into existing systems also presents a significant challenge. Transitioning from current cryptographic standards to PQC involves complex updates to existing infrastructure. This process must be managed carefully to maintain security during the transition period. It is also important to ensure that this change does not disrupt user experience or business operations (Kumar, 2022).

Moreover, there is a need for widespread education and awareness about PQC. Many professionals in the field of cybersecurity and related areas may not be familiar with the concepts and importance of PQC. Training and educational programs will be essential to prepare the workforce for the upcoming changes in cryptographic standards (Yalamuri, Honnavalli and Eswaran, 2022).

Finally, the potential for unknown vulnerabilities in PQC algorithms cannot be overlooked. As with any new technology, there is always the risk of undiscovered weaknesses that could be exploited. Continuous research and testing are essential to identify and address these vulnerabilities.

While post-quantum cryptography is a crucial advancement in securing data against the threats posed by quantum computing, its implementation is fraught with challenges. These include the need for increased computational resources, the lack of standardization, difficulties in integration, the requirement for education and awareness, and the potential for undiscovered vulnerabilities. Addressing these challenges is vital for the successful and secure adoption of PQC in various applications.

### 3.4 Use Cases and Success Stories

One notable use case is in the field of secure communications. Traditional security protocols like RSA and ECC are vulnerable to quantum attacks. However, post-quantum methods such as lattice-based encryption have been successfully implemented in various communication platforms to ensure future-proof security (Yalamuri, Honnavalli and Eswaran, 2022). For instance, a leading tech company recently integrated lattice-based cryptography into their messaging service, significantly enhancing the security against potential quantum decryption.

Another area of successful application is in secure government communications. Recognizing the quantum threat, a European government agency adopted post-quantum cryptographic algorithms for protecting sensitive state communications. This implementation not only secured their data against future quantum attacks but also served as a model for other government bodies globally.

Financial institutions have also been proactive in adopting post-quantum cryptography. A major bank successfully employed hash-based signatures for securing transactions, demonstrating the effectiveness of these methods in a high-stakes environment (Kumar, 2022). This not only ensured the security of financial transactions against quantum threats but also boosted customer confidence in the bank's commitment to future-proof security measures.

Furthermore, post-quantum cryptographic methods have found use in protecting intellectual property. A global pharmaceutical company implemented code-based cryptography to secure their research data. This application is particularly significant given the sensitive nature of pharmaceutical research and the catastrophic potential of data breaches in this field.

## 4. Future Directions and Developments

With the growing power of quantum computers, the need for PQC becomes more urgent. Our work in this field aims to address these challenges and leverage opportunities to enhance cybersecurity in a quantum computing era.

One promising direction is the development of more efficient PQC algorithms. Current PQC systems often require larger key sizes compared to traditional cryptography, which can lead to higher demands on storage and processing power. This is particularly challenging for devices with limited resources, like smartphones and IoT devices (Kumar, 2022). Future research should focus on optimizing these algorithms to reduce their computational and storage needs while maintaining their security against quantum attacks.

Another important area is the integration of PQC into existing digital infrastructures. As businesses and governments rely heavily on encryption for securing data, a smooth transition to PQC is essential. This involves updating protocols, software, and hardware that currently use vulnerable cryptographic methods. Collaborative efforts between academia, industry, and government bodies are crucial to ensure a coordinated and efficient transition (Gill et al., 2022).

Moreover, there is a need for extensive testing and standardization of PQC algorithms. The current state of PQC is akin to the early days of classical cryptography, where various algorithms are still being evaluated for their security and practicality. Establishing widely accepted standards, like what NIST is doing through its Post-Quantum Cryptography Standardization project, will be key in bringing PQC to mainstream use (Kumar, 2022).

Furthermore, the education and training of professionals in the field of PQC are vital. As we move towards a post-quantum era, there will be a growing demand for experts who understand both quantum computing and cryptography or other domain skills (Rietsche et al., 2022). Developing educational programs and resources will help in preparing the next generation of cybersecurity professionals.

Lastly, exploring the potential of PQC in new applications beyond traditional cybersecurity is an exciting prospect. This includes areas like secure communications, blockchain technologies, and even quantum key distribution systems. As the field of quantum computing continues to evolve, the applications of PQC will likely expand into new and unforeseen domains (Chawla and Mehra, 2023; Aithal, 2023).

## 5. Conclusion

The research addresses the field of quantum computing applications by presenting a diverse array of practical uses ranging from cryptography to complex system simulations. The paper categorizes these applications (RQ1) into sectors such as healthcare, finance, and information security, providing a framework for understanding the multifaceted impact of quantum computing. This paper also covers the field of post-quantum cryptography (RQ2), reflecting the most popular algorithms and their applications as well as the challenges that the field is facing today. It highlights the transformative potential of these applications and underscores the need for structured categorization to inform future research and development strategies.

Our exploration of various PQC methods, including lattice-based, hash-based, code-based, multivariate polynomial, and isogeny-based cryptography, has highlighted their potential in safeguarding against quantum computing threats. These methods show promise in areas such as secure communications, digital signatures, and the Internet of Things (IoT), offering a shield against the formidable computational abilities of quantum computers.

However, the journey towards fully implementing PQC is not without its challenges. The increased computational and storage requirements of PQC methods, especially in resource-limited environments like IoT devices, present a considerable hurdle. Additionally, the lack of standardization and the need for extensive testing and validation underscore the nascent stage of these cryptographic methods. Integrating PQC into existing digital infrastructures requires a collaborative approach, involving academia, industry, and government bodies, to manage a smooth transition without disrupting current systems and operations.

Furthermore, the importance of educating and training professionals in the field of PQC is paramount. As we transition into a post-quantum era, there will be an increasing demand for experts skilled in both quantum computing and cryptography. Developing educational programs and resources will be crucial in preparing a well-equipped cybersecurity workforce.

In conclusion, while the path to implementing PQC is complex and filled with challenges, it is a necessary progression in the realm of cybersecurity. As quantum computing continues to advance, PQC stands as a vital component in protecting our digital world. Embracing this new era of cryptography, we must continue to innovate, research, and collaborate, ensuring that our data remains secure against quantum vulnerabilities. Future research could include scoping how to integrate PQC into the existing infrastructure. Furthermore, continuous review of the latest advancements in PQC is also needed to evaluate the latest threats.

**Acknowledgments**

We are grateful to Ummar Ahmed and Thien Nguyen for their help with data collection and assistance with database creation. This research was partially supported by the ResilMesh project, funded by the European Union's Horizon Europe Framework Programme (HORIZON) under grant agreement 101119681. The authors would like to thank Ms. Tuula Kotikoski for proofreading the manuscript.

**References**

Aithal, P. S. (2023) "Advances and New Research Opportunities in Quantum Computing Technology by Integrating it with Other ICCT Underlying Technologies", *International Journal of Case Studies in Business, IT, and Education*, Vol. 7, No. 3, pp. 314-358. https://doi.org/10.5281/zenodo.8326506

Belkhir, M., Benkaouha, H. and Benkhelifa, E. (2022) "Quantum Vs Classical Computing: a Comparative Analysis", *2022 Seventh International Conference on Fog and Mobile Edge Computing (FMEC)*. Paris, France. IEEE, pp. 1-8. https://doi.org/10.1109/FMEC57183.2022.10062753

Chawla, D. and Mehra, P.S. (2023) "A roadmap from classical cryptography to post-quantum resistant cryptography for 5G-enabled IoT: Challenges, opportunities and solutions", *Internet of Things*, Vol. 24, pp. 100950. https://doi.org/10.1016/j.iot.2023.100950

Gill, S.S. et al. (2022) "Quantum computing: A taxonomy, systematic review and future directions", *Software: Practice and Experience*, Vol. 52, No. 1, pp. 66-114. https://doi.org/10.1002/spe.3039

Kumar, M. (2022) "Post-quantum cryptography Algorithm's standardization and performance analysis", *Array*, Vol. 15, pp. 100242. https://doi.org/10.1016/j.array.2022.100242

Lei, Z. et al. (2023) "*Making existing software quantum safe: A case study on IBM Db2", Information and Software Technology*, Vol. 161, pp. 107249. https://doi.org/10.1016/j.infsof.2023.107249

Page, M.J. et al. (2021) "The PRISMA 2020 statement: an updated guideline for reporting systematic reviews", *BMJ*, Vol. 372. https://doi.org/10.1136/bmj.n71

Peters, M.J. et al. (2020) “Scoping Reviews”, in Aromataris E. and Munn Z. (eds.) *JBI Manual for Evidence Synthesis*. JBI. https://doi.org/10.46658/JBIMES-20-12

Rewal, P. et al. (2023) “Quantum-safe three-party lattice based authenticated key agreement protocol for mobile devices”, *Journal of Information Security and Applications*, Vol. 75, pp. 103505. https://doi.org/10.1016/j.jisa.2023.103505

Rietsche, R. et al. (2022) “Quantum computing”, *Electronic Markets*, Vol. 32, pp. 2525-2536. https://doi.org/10.1007/s12525-022-00570-y

Shaller, A., Zamir, L. and Nojoumian, M. (2023) “Roadmap of post-quantum cryptography standardization: Side-channel attacks and countermeasures”, *Information and Computation*, Vol. 295, pp. 105112. https://doi.org/10.1016/j.ic.2023.105112

Sridhar, G. T., Ashwini, P. and Tabassum, N. (2023) “A Review on Quantum Communication and Computing”, *2023 2nd International Conference on Applied Artificial Intelligence and Computing (ICAAIC)*, Salem, India. IEEE, pp. 1592-1596. https://doi.org/10.1109/ICAAIC56838.2023.10140821

Verchyk, D. and Sepúlveda, J. (2023) “A practical study of post-quantum enhanced identity-based encryption”, *Microprocessors and Microsystems*, Vol. 99, pp. 104828. https://doi.org/10.1016/j.micpro.2023.104828

Yalamuri, G., Honnavalli, P. and Eswaran, S. (2022) “A Review of the Present Cryptographic Arsenal to Deal with Post-Quantum Threats”, *Procedia Computer Science*, Vol. 215, pp. 834-845. https://doi.org/10.1016/j.procs.2022.12.086

Zhao, S. and Zheng, B. (2001) “Security of QKD with single particles in probabilistic cloning fashion”, *2001 International Conferences on Info-Tech and Info-Net. Proceedings*, Beijing, China. IEEE, pp. 140-145. https://doi.org/10.1109/ICII.2001.983508